\documentclass{article}
\usepackage{spconf,amsmath,graphicx, amssymb}
\usepackage{algorithmic}
\usepackage[caption=false,font=footnotesize]{subfig}
\usepackage[noadjust]{cite}
\usepackage{setspace}

\newsavebox{\ieeealgbox}
\newenvironment{boxedalgorithmic}
{\begin{lrbox}{\ieeealgbox}
    \begin{minipage}{\dimexpr\columnwidth-2\fboxsep-2\fboxrule}
      \begin{algorithmic}}
      {\end{algorithmic}
    \end{minipage}
  \end{lrbox}\noindent\fbox{\usebox{\ieeealgbox}}
}

\title{Unit circle MVDR beamformer}

\name{Saurav R. Tuladhar, John R. Buck \sthanks{Supported by ONR grant N00014-12-1-0047.}}
\address{University of Massachusetts Dartmouth\\
	Electrical and Computer Engineering Department\\
	North Dartmouth, Massachusetts, USA}

\begin{document}
\ninept
%\onecolumn
%\doublespacing
\maketitle

\begin{abstract}
  The array polynomial is the z-transform of the array weights for a
  narrowband planewave beamformer using a uniform linear array
  (ULA). Evaluating the array polynomial on the unit circle in the
  complex plane yields the beampattern. The locations of the
  polynomial zeros on the unit circle indicate the nulls of the
  beampattern. For planewave signals measured with a ULA, the
  locations of the ensemble MVDR polynomial zeros are constrained on
  the unit circle. However, sample matrix inversion (SMI) MVDR
  polynomial zeros generally do not fall on the unit circle. The
  proposed unit circle MVDR (UC MVDR) projects the zeros of the SMI
  MVDR polynomial radially on the unit circle. This satisfies the
  constraint on the zeros of ensemble MVDR polynomial. Numerical
  simulations show that the UC MVDR beamformer suppresses interferers
  better than the SMI MVDR and the diagonal loaded MVDR beamformer and
  also improves the white noise gain (WNG).
\end{abstract}
%
%Command shortcuts for various math symbols
% Inherited from JRB and KEW
\newcommand{\Cov}{\boldsymbol{\Sigma}}
\newcommand{\cov}{\sigma}
\newcommand{\Covdmr}{\Cov_{\rm DMR}}
\newcommand{\eval}{\gamma}
\newcommand{\Eval}{\boldsymbol{\Gamma}}
\newcommand{\evec}{\boldsymbol{\xi}}
\newcommand{\Evec}{\boldsymbol{\Xi}}
\newcommand{\sampCov}{{\bf S}}
\newcommand{\sampcov}{s}
\newcommand{\sampCovdmr}{\sampCov_{\rm DMR}}
\newcommand{\sampeval}{g}
\newcommand{\sampEval}{{\bf G}}
\newcommand{\sampevec}{{\bf e}}
\newcommand{\sampEvec}{{\bf E}}
\newcommand{\herm}{^{\rm H}}
\newcommand{\rep}{{\bf v}}        % planewave replica
\newcommand{\repmat}{{\bf V}}     % replica matrix
\newcommand{\sigamp}{b}
\newcommand{\wmvdr}{{\bf w}_{\rm MVDR}}
\newcommand{\dl}{\delta}
\newcommand{\wconv}{{\bf w}_{\rm conv}}
\newcommand{\nulldepth}{\mbox{ND}}
\newcommand{\replook}{\rep_0}
\newcommand{\repint}{\rep_1}
\newcommand{\tsp}{^{\rm T}}
\newcommand{\inv}{^{-1}}
\newcommand{\limrmt}{\lim_{RMT}}
\newcommand{\eig}{\operatorname{eig}}
\newcommand{\diag}{\operatorname{diag}}
\newcommand{\sign}{{\operatorname{sgn}}}

% SRT new commands
\newcommand{\ulook}{u_0}
\newcommand{\uinter}{u_1}
\newcommand{\beampatu}{\text{B}(u)}
\newcommand{\beampat}[1]{\text{B}(#1)}
\newcommand{\wcbf}{{\bf w}_{\rm CBF}}
\newcommand{\wsmi}{{\bf w}_{\rm SMI}}
\newcommand{\wuc}{{\bf w}_{\rm UC}}
\newcommand{\wt}{{\bf w}}
\newcommand{\beampolyz}{\mbox{P}(z)}
\newcommand{\mvdrpoly}{P_M(z)}
\newcommand{\cbfpoly}{P_C(z)}
\newcommand{\smipoly}{P_S(z)}
\newcommand{\ucpoly}{P_{UC}(z)}
\newcommand{\datavec}{{\bf x}}
\newcommand{\noisevec}{{\bf n}}
\newcommand{\zvec}{\boldsymbol{z}}
\newcommand{\ztrans}{\mathcal{Z}}
\newcommand{\eye}{{\bf I}}     % replica matrix
\newcommand{\srt}[1]{\textbf{SRT:#1}}
\newcommand{\uc}{\mathcal{U}}
\newcommand{\cnormal}[2]{\mathcal{CN}(#1  #2)}
\newcommand{\sampz}{\xi}
\newcommand{\ensz}{\zeta}
\newcommand{\ucz}{\hat{\xi}}
\newcommand{\zinter}{\ensz{}_I}
\newcommand{\pout}{P_o}
\newcommand{\notchdepth}{\text{ND}}

% ---------- Introduction -------------

\section{Introduction}
\label{sec:intro}
Beamformers enhance signals arriving at an array from a desired look
direction while suppressing interferers and noise. Conventional
beamformers (CBFs) using a delay-and-sum approach have limited ability
to suppress loud interferers which can leak through the high sidelobes
of the CBF beampattern and mask weaker signal of interest. Adaptive
beamformers (ABF) place notches in the direction of interferers to
suppress the interferer power at the output and improve
signal-to-interferer-plus-noise ratio (SINR) \cite{vtree2002oap}. The
minimum variance distortionless response (MVDR) beamformer is one of
the most commonly used ABFs \cite{capon1969mvdr}. In practice, the
ensemble covariance matrix (ECM) is unknown so the sample covariance
matrix (SCM) replaces the ECM to compute the ABF weights. The
resulting ABF is known as the Sample Matrix Inversion (SMI) MVDR
beamformer \cite{vtree2002oap}.

The beampattern defines the spatial response of a beamformer
\cite{vtree2002oap}. The beampattern of a beamformer using a ULA can
be represented as an array polynomial by taking z-transform of
beamformer weights \cite{Schelkunoff1943array}. This is analogous to
the system function representation of a discrete time (DT) LTI filter
by taking z-transform of its impulse response \cite{Oppenheim1989}. As with DT LTI filters,
beamformers also have a pole-zero representation in the complex plane
and continuing the analogy, the beampattern is obtained by evaluating
the array polynomial on the unit circle. The array polynomial zeros
generally correspond to the beampattern notches but when the zeros
fall on the unit circle they result in perfect notches or nulls.

The zeros of an ensemble MVDR beamformer polynomial for narrowband
planewave are constrained on the unit circle. However, the SMI MVDR
array polynomial zeros generally do not lie on the unit circle. A new
beamformer is developed by radially projecting the SMI MVDR zeros on
the unit circle and satisfying the constraint on the ensemble MVDR
zeros. The proposed unit circle MVDR (UC MVDR) beamformer suppresses
interferers better than the SMI MVDR and the diagonal loaded MVDR
beamformer and at the same time improves white noise gain (WNG)
performance.

Prior work involving the use of array polynomial representation for
beamforming includes work by
Steinberg \cite{Steinberg1976}. Steinberg discusses the polynomial
representation of radiation pattern of uniform antenna arrays and
presents an approach to synthesize radiation patterns by manipulating
zero locations on the unit circle. The methods presented in
\cite{Steinberg1976} are limited to ensemble cases. Several proposed
adaptive notch filters for DT signals constrain the filter poles and
zeros to render 'sharper' notches in their frequency response
\cite{Nehorai1985, Friedlander1984notch, Shynk1986}. However these
approaches are based on DT IIR filters while beamformers with ULAs are
analogous to DT FIR filters.

The remainder of this paper is organized as follows:
Sec.~\ref{sec:background} reviews the signal model, MVDR beamformer and the
metrics used to evaluate beamformer performance. Sec.~\ref{sec:array-poly}
develops the polynomial representation for ULA beamformers and discusses
zero locations for the MVDR and SMI MVDR
beamformers. Sec.~\ref{sec:ucmvdr} presents the UC MVDR beamformer
algorithm. Sec.~\ref{sec:results} discusses the simulation results comparing
the performance of the UC MVDR with SMI MVDR and DL MVDR.

% ---------- Background --------------
\section{Background}
\label{sec:background}
The narrowband planewave data measured on an $N$ element ULA is modeled
as an $N \times 1$ complex vector,
\begin{equation}
  \label{eq:array-data} \datavec = \sum\limits_{i=1}^D a_i\rep_i +
\noisevec
\end{equation}
where $D$ is the number of planewave signals, $a_i$ is $i^{th}$ signal
amplitude and $\noisevec$ is the noise sample vector. The amplitude is
modeled as a zero mean complex circular Gaussian random variable,
i.e., $a_i \sim \cnormal{0, \cov_i^2}$ and the noise is assumed to be
spatially white with complex circular Gaussian distribution, i.e.,
$\noisevec \sim \cnormal{\mathbf{0}, \cov_w^2\eye}$. The complex
vector $\rep_i$ is the narrowband planewave array manifold vector
defined as,
\[
  \rep_i = [1,~e^{-j{(2\pi/\lambda)} d u_i},~ e^{-j{(2\pi/\lambda)}2d u_i},
\ldots, e^{-j{(2\pi/\lambda)}(N-1) d u_i}]^\text{T},
\]
where $u_i = \cos(\theta_i)$ and $\theta_i$ is the $i^{th}$ signal
direction, $\lambda$ is the wavelength, $d$ is the ULA inter element
spacing and $[\cdot]^T$ denotes transpose. In the sequel, signal
direction will be represented in terms of $u$. Assuming the $D$
signals to be uncorrelated in \eqref{eq:array-data}, the ECM is,
\begin{align}
  \label{eq:ecm} \Cov =& E[\datavec\datavec\herm] = \sum\limits_{i =
1}^D\cov_i^2\rep_i\rep_i\herm + \cov_w^2\eye. % =& \repmat S
\end{align}
where $\cov_i^2$ is $i^{th}$ signal power and $\cov_w^2$ is the noise power at
each sensor. 

The narrowband planewave MVDR beamformer weight vector for a ULA is,
\begin{equation}
  \label{eq:mvdr-wt} \wmvdr =
\frac{\Cov\inv\replook}{\replook\herm\Cov\inv\replook},
\end{equation}
where $\rep_0$ is the array manifold vector for the look direction $\ulook
= \cos(\theta_0)$. Computing the MVDR weights in \eqref{eq:mvdr-wt}
requires knowledge of the ECM but in practical applications the ECM is unknown
\textit{a priori}. Consequently, the MVDR ABF is approximated by the SMI
MVDR computed by replacing the ECM $\Cov$ in \eqref{eq:mvdr-wt} with the SCM,
\[
  \sampCov = \frac{1}{L}\sum\limits_{l=1}^{L}\datavec_l\datavec_l\herm  
\]
where the $L$ is the number of data snapshots and $\datavec_l$ is the
data snapshot vector in \eqref{eq:array-data}. In practice, the number
of data snapshots available to compute the SCM are limited either due
to physical or stationarity constraints of the environment
\cite{riba1997comm}\cite{baggeroer1999passive}. When the number of
snapshots is on the order of the number of array elements, i.e.,
$N \approx L$, the SCM is ill-conditioned for inversion. The resulting
SMI MVDR beamformer suffers from a distorted beampattern with high
sidelobes and subsequent loss in SINR
\cite{reed1974rapid}\cite{Carlson1988scm}. In this scenario, a common
approach is to apply diagonal loading (DL) to the SCM to get
$\sampCov_\dl = \sampCov + \dl\eye{}$ where $\dl$ is the DL
factor. This results in the DL MVDR beamformer. DL makes SCM inversion
stable, provides better sidelobe control and improves beamformer WNG
\cite{mestre2005diagonal}. An appropriate choice for the DL factor
requires knowledge of the signal and the interference plus noise power
levels which are unknown in practice. Some of the proposed methods to
choose DL factor are either ad-hoc in nature or require numerical
solutions to optimization problems \cite{mestre2005diagonal}.

The beampattern defines the complex gain due to the
beamformer on a unit amplitude planewave from direction $u =
\cos(\theta)$, i.e.,
\begin{equation}
  \label{eq:beampatu} 
  \beampatu = \wt^H\rep(u) = \sum_{n=0}^{N-1}w_n^*\left(e^{-j\frac{2\pi}{\lambda} d u}\right)^n
\end{equation}
where $(\cdot)^*$ denotes conjugate and $-1 \leq u \leq 1$ is the direction
range of the beampattern. In the presence of strong interfering signals, a
beamformer's ability to suppress the interferers is quantified by the notch
depth (ND) defined as $\notchdepth = |\beampat{\uinter}|^2$, where
$\uinter$ is the interferer direction. ABFs aim to improve SINR by
adjusting the ND and location based on interferer power and direction.

White noise gain (WNG) is defined as the array gain when the noise is
spatially white. Assuming unity gain in the look direction,
$\text{WNG} = ||\wt||^{-2}$ where $||\cdot||$ denotes the Euclidean
norm \cite{vtree2002oap}. WNG is also a metric for beamformer
robustness against mismatch \cite{Gilbert1955}. The CBF has the
optimal WNG which is equal to the number of array sensors $N$
\cite{Cox1987robust}.

% --------- UCMVDR
\section{Beamformer Polynomial}
\label{sec:array-poly}
The beampattern of a narrowband planewave beamformer with ULA can be
represented as a complex polynomial
\cite{Schelkunoff1943array}\cite{Steinberg1976}. For a standard ULA
with $d = \lambda/2$,
\begin{equation}
  \label{eq:beampat-ula}
  \text{B}(u) = \sum\limits_{n=0}^{N-1} w^*_n( e^{-j\pi u})^n.
\end{equation}
Letting $z = e^{j\pi u}$ in \eqref{eq:beampat-ula}, we get the array polynomial
\begin{equation}
  \label{eq:beampat-poly}
  \beampolyz = \sum\limits_{n=0}^{N-1} w^*_n z^{-n} = \ztrans(\wt\herm).
\end{equation}
$\beampolyz$ is an $N-1$ degree polynomial in the complex variable $z$
with beamformer weights ($w^*_n$) as its
coefficients. Eq. \eqref{eq:beampat-poly} is in the form of the
z-transform of the conjugate beamformer weights \cite[Chap.3]{Oppenheim1989}. This polynomial representation maps the
bearing variable $u$ into the complex plane. The phase of the complex
variable is related to the bearing variable as $\operatorname{arg}(z)
= \omega = \pi u$. Evaluating \eqref{eq:beampat-poly} on the unit
circle $\lbrace z \in \mathbb{C}, |z| = 1\rbrace$ returns it to
\eqref{eq:beampat-ula}. Hence the zeros of $\beampolyz$ on the unit
circle correspond to nulls of the beampattern.

% A CBF with look direction at broadside ($\ulook = 0$) has weight
% vector $\wcbf = 1/N$. The CBF polynomial can be derived by using
% $\wcbf$ in \eqref{eq:beampat-poly} to get
% \begin{align}
%   \label{eq:cbf-poly}
%   \cbfpoly = \frac{1}{N}\sum\limits_{n=0}^{N-1}z^{-n} 
%   = \frac{1}{N}\left[\frac{z^N - 1}{z^{N-1}(z - 1)}\right].
% \end{align}
% The zeros of \eqref{eq:cbf-poly} are the $N^{th}$ roots of unity
% except at $z = 1$. When the CBF is steered to a different look
% direction, the CBF zeros merely rotate along the unit circle
% \cite{Steinberg1976}. Hence all CBF zeros lie on $\uc$.

\begin{figure}[t]
\centering
\includegraphics[width=2.6in]{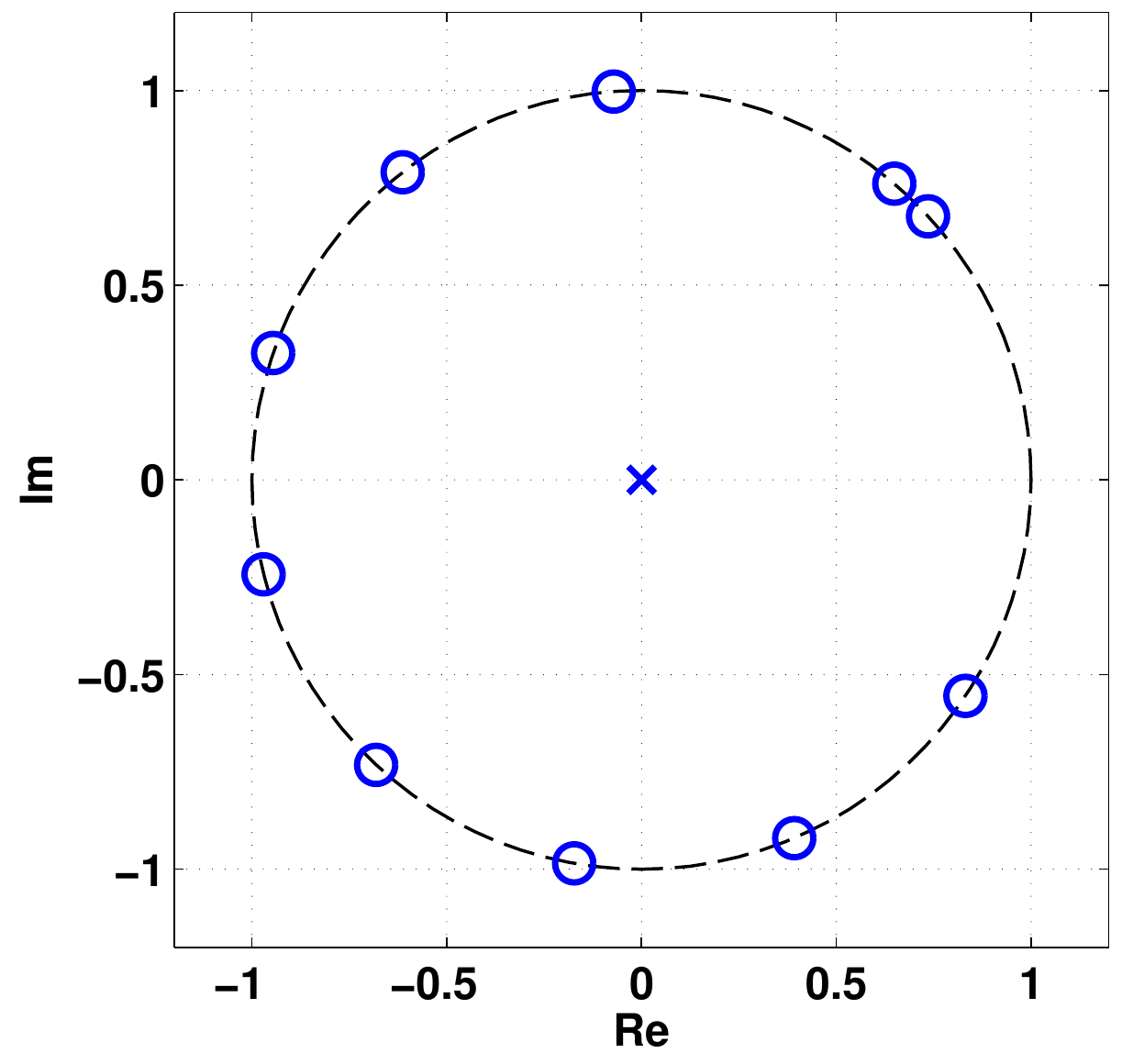} 
\caption{Zero locations of MVDR beamformer using $N= 11$ element ULA.}
 \label{fig:cbf-mvdr-pzplot}
\end{figure}

% \subsection{MVDR polynomial}
% \label{sec:mvdr-poly}
The MVDR polynomial is obtained as
\begin{equation}
  \label{eq:mvdr-poly}
  \mvdrpoly = \ztrans(\wmvdr\herm) =  \Gamma \prod\limits_{n=1}^{N-1}(1 - \ensz_n z\inv),
\end{equation}
where $\Gamma$ is a scaling term and $\ensz_n$ are the ensemble MVDR
zeros. \figurename{} \ref{fig:cbf-mvdr-pzplot} shows MVDR zeros for an
example case of $N = 11$ element ULA and a single interferer at
$\uinter = 3/N$. All MVDR zeros in \figurename{}
\ref{fig:cbf-mvdr-pzplot} are on the unit circle. In fact the MVDR
ensemble zeros are always constrained on the unit circle for planewave
beamforming using a ULA. The unit circle constraint was initially
discovered and proved by Seinhardt and Guerci but the result does not
appear to be widely known \cite{steinhardt2004stap}.

However, the SMI MVDR zeros are perturbed from the ensemble MVDR zero
locations and are randomly located on the complex plane about the
ensemble MVDR zero locations. The SMI MVDR zeros do not necessarily
lie on the unit circle and they correspond to notches in the SMI MVDR
beampattern. Any zeros that fall closer to the origin or far outside
the unit circle have negligible contribution to beampattern
\cite[Chap.~5]{Oppenheim1989}. The following section describes how the
SMI MVDR beamformer can be modified by moving the sample zeros to the
unit circle following the constraint on ensemble zeros.

\section{Unit circle MVDR beamformer}
\label{sec:ucmvdr}
The unit circle MVDR (UC MVDR) beamformer projects the SMI MVDR zeros
radially on the unit circle consistent with the constraint on
the ensemble MVDR zeros. By placing zeros on the unit circle, the UC
MVDR beampattern guarantees nulls in the direction corresponding to
the zeros. \figurename{} \ref{algo-ucmvdr} describes the UC MVDR
beamformer algorithm. The algorithm begins from the SMI MVDR weights
$\wsmi$ computed using the SCM. The z-transform of the elements of
$\wsmi\herm$ gives the SMI MVDR polynomial
\[
  \smipoly = G \prod\limits_{n=1}^{N-1}(1 - \sampz_n z\inv),  
\]
where $G$ is a scaling factor, $\sampz_n = r_ne^{j\omega_n}$ are the
SMI MVDR zeros and the $r_n$s are generally not unity. Each SMI MVDR
zero $\sampz_n$ is moved radially to the unit circle to obtain the UC
MVDR zeros $\ucz_n = e^{j\omega_n}$. An exception is made when the SMI
MVDR zeros fall within the CBF main-lobe region in the complex
plane. Such zeros are moved to the CBF first-null location on the unit
circle to protect the main-lobe. A unit circle polynomial $\ucpoly$ is
defined using the new unit circle zeros $\ucz_n$s,
\begin{align}
  \label{eq:uc-poly}
  \ucpoly =  \prod\limits_{n=1}^{N-1}(1 - \ucz_n z\inv) = \sum\limits_{n=0}^{N-1} c_n^* z\inv.
\end{align}
The coefficients $c_n$s are the new beamformer weights. The resulting
beamformer will have beampattern nulls in the direction corresponding
to $\ucz_n$. Finally, the $c_n$s are scaled to ensure the beamformer
has unity gain in the look direction to satisfy the distortionless
constraint of a MVDR beamformer. The resulting UC MVDR beamformer
weight vector is $\wuc = {\bf{c}}/{|\bf{c}\herm\replook|}$ where
$\mathbf{c} = [c_0, c_2 \ldots c_{N-1}]$ and $\replook$ is the array
manifold vector for look direction $\ulook = \cos(\theta_0)$. Since
polynomial zero locations are invariant to coefficient scaling, the UC
MVDR beampattern will still have nulls in the same locations as
$\ucpoly$.

\begin{figure}[!th]
\centering
\subfloat[]{\includegraphics[width=2.7in]{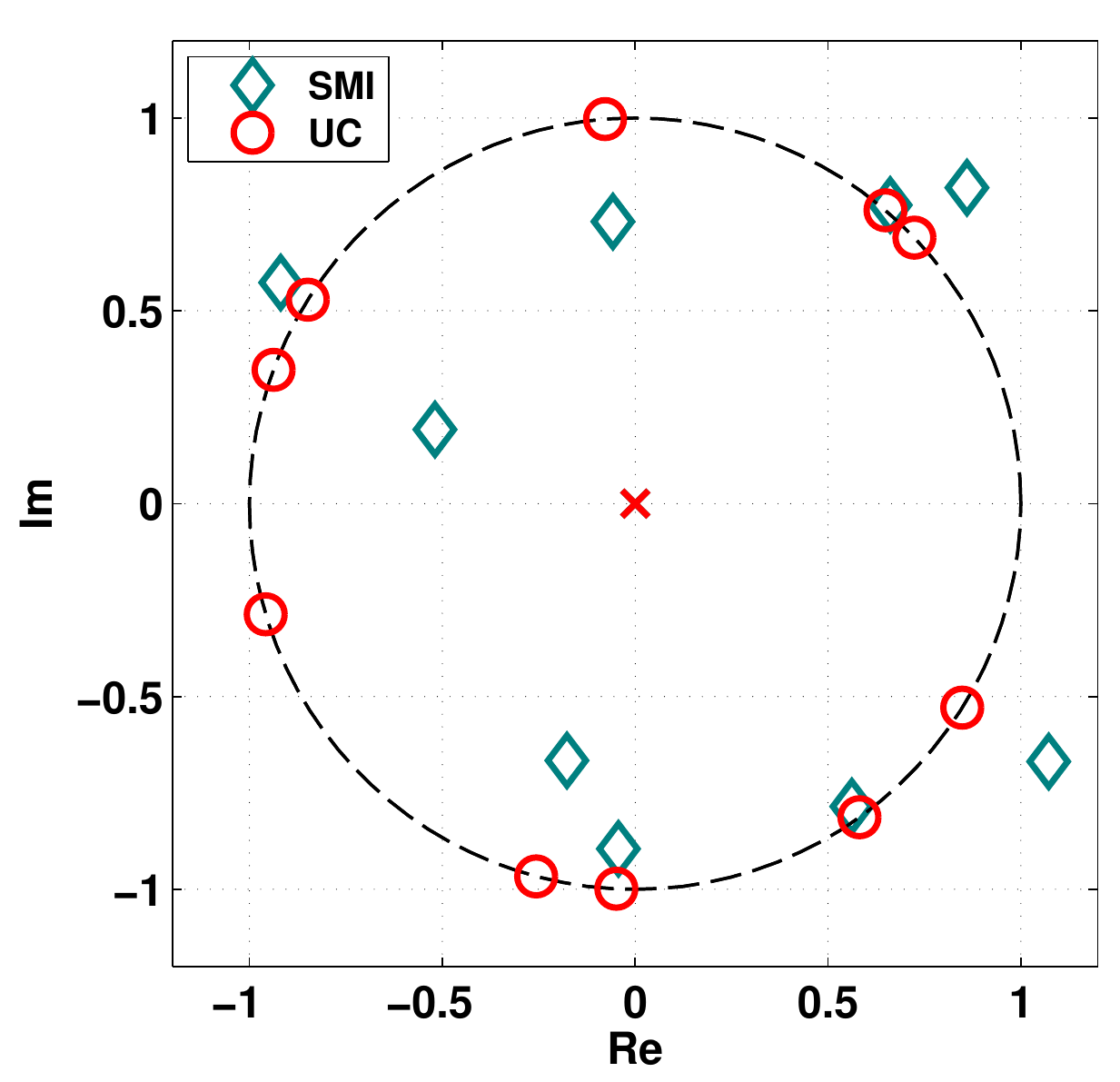}  \label{fig:smi-ucmvdr-pzplot}}
\hfill
\subfloat[]{\includegraphics[width=2.7in]{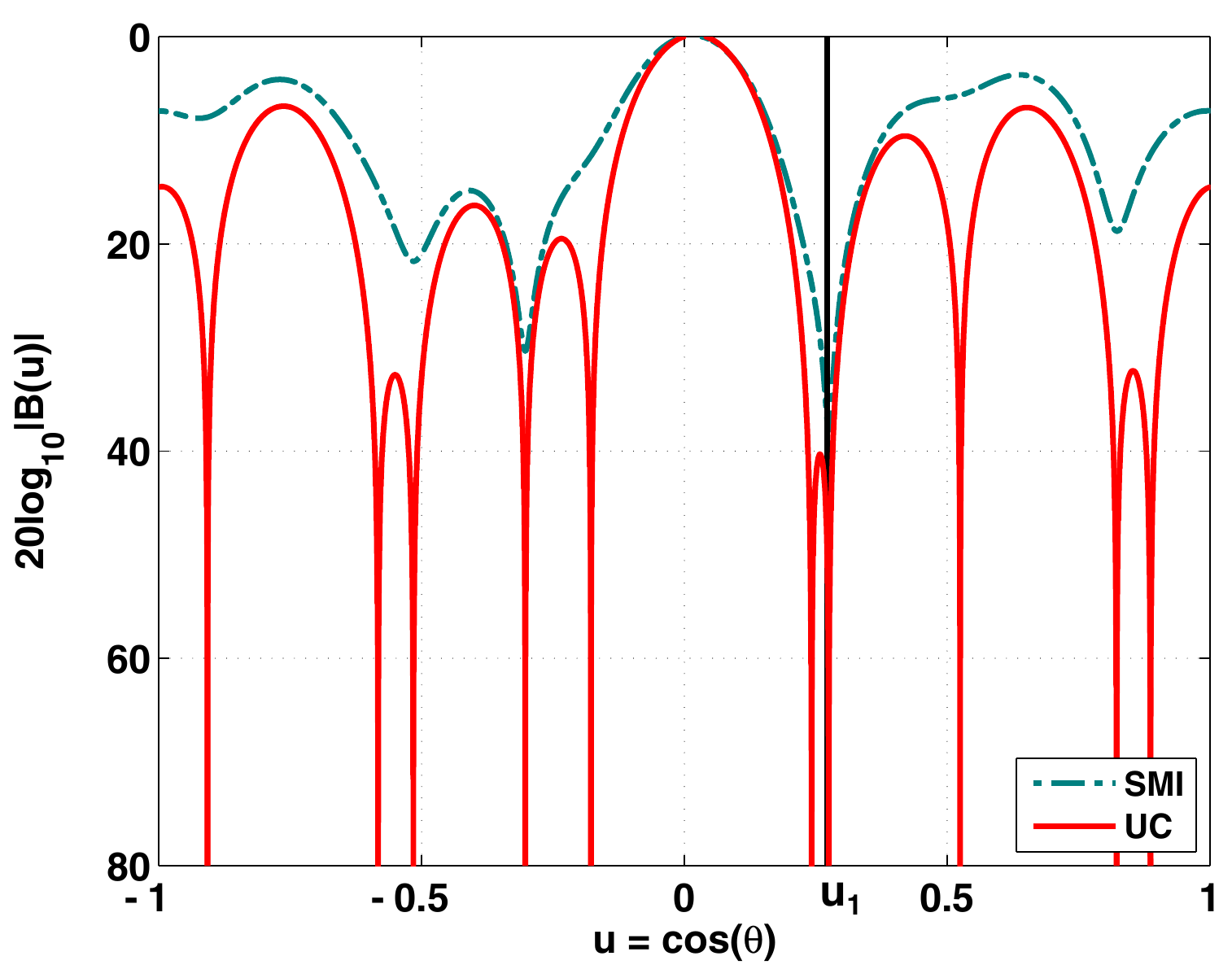}  \label{fig:smi-ucmvdr-bpplot}}
\caption{Zero locations and beampattern for representative example of
  SMI MVDR and UC MVDR beamformer using $N = 11$ element ULA for $L = 11$ snapshots.}
\label{fig:smi-ucmvdr-plots}
\end{figure}

\figurename{} \ref{fig:smi-ucmvdr-plots} shows a representative
example of zero locations and beampattern of a UC MVDR compared with
SMI MVDR beamformer using $N = 11$ element ULA and $L = 12$
snapshots. A single interferer is present at $\uinter = \cos(\theta) =
3/N$. In \figurename{} \ref{fig:smi-ucmvdr-pzplot}, the green diamond
markers indicate the SMI MVDR zero locations and the red circle
markers indicate the UC MVDR zeros obtained by moving the SMI MVDR
zeros to unit circle. The corresponding beampattern plots in
\figurename{} \ref{fig:smi-ucmvdr-bpplot} show perfect notches and
lower sidelobes in the UC MVDR beampattern (solid red) in contrast to
shallow notches and higher sidelobes in the SMI MVDR beampattern (dot-dash
green).

\begin{figure}[t]
  \begin{boxedalgorithmic}[1]    
    \STATE Compute SCM : $\sampCov =
    \frac{1}{L}\sum\limits_{n=1}^N\datavec\datavec\herm$
     \STATE Compute SMI MVDR weights : $\wsmi =
     {\sampCov\inv\replook}/{(\replook\sampCov\inv\replook)}$
     \STATE $\smipoly = \ztrans (\wsmi\herm) = G\prod\limits_{n=1}^{N-1}(1 -
     \sampz_nz\inv)$ and  $\sampz_n = r_ne^{j\omega_n}$
%     \STATE Move zeros $\sampz_n$ to unit circle 
     \IF{$|\omega_n| > 2\pi/N$}
     \STATE $\ucz_n = e^{j\omega_n}$     
     \ELSIF{$|\omega_n| \leq 2\pi/N$}
     \STATE $\ucz_n = e^{j\sign(\omega_n){2\pi/N}}$
     \ENDIF
     \STATE Use $\ucz_n$ to create new unit circle polynomial  : \\
     $\ucpoly = \prod\limits_{n=1}^{N-1}(1 - \ucz_n z\inv) =
        \sum\limits_{n=0}^{N-1} c_n^*z^{-n}$ 
     \STATE Define : $\mathbf{c} = [c_1, c_2 \ldots c_N]$
     \STATE UC MVDR weight : $\wuc = \mathbf{c}/|\mathbf{c}\herm\replook|$
  \end{boxedalgorithmic}
  \caption{UC MVDR beamformer algorithm}
  \label{algo-ucmvdr}

\end{figure}

% ----------- results
\section{Simulation Results}
\label{sec:results}
% assumed. The results were obtained from a 5000 trial Monte Carlo
% experiments.The results show that the UC-MVDR achieves better output
% power suppression and higher WNG compared to SMI MVDR.

\begin{figure}[!t]
\centering
\includegraphics[width=3in]{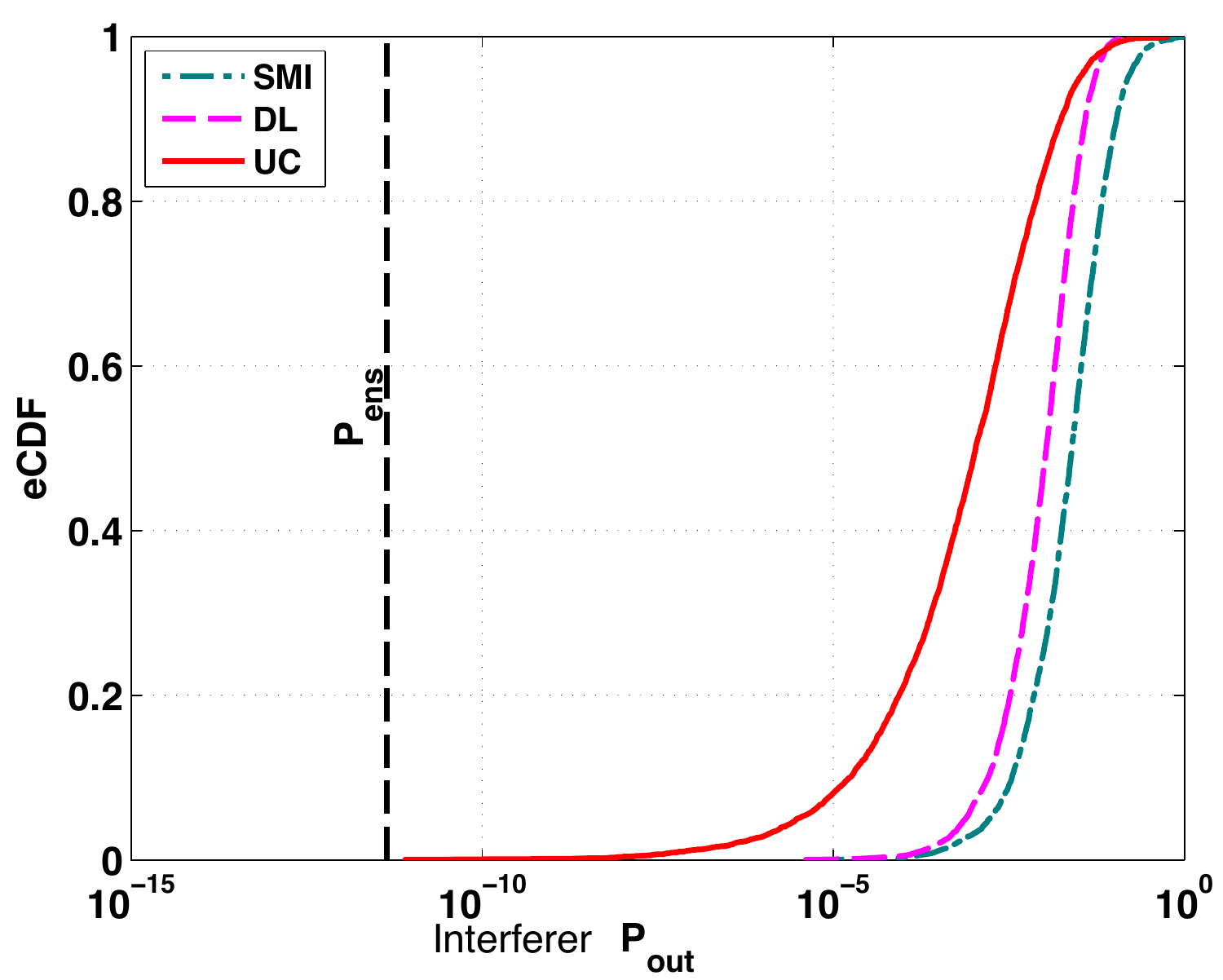}
\caption{Comparison of interferer output power empirical CDF between UC
  MVDR, SMI MVDR and DL MVDR beamformers with $N = 11$ element ULA and $L =
  12$ snapshots.}
\label{fig:ecdf-plots}
\end{figure}

\figurename{} \ref{fig:ecdf-plots} compares the empirical CDF of
output power in the interferer direction for the UC MVDR beamformer
against the SMI MVDR and DL MVDR beamformers. The CDF curves are based on
5000 Monte Carlo trials. The dashed vertical line represents the ideal
output power using the ensemble MVDR beamformer. The ULA size was $N =
11$ and a single interferer was fixed at $\uinter = \cos(\theta_1) =
3/N$ for each trial. The sensor level INR was 40 dB. The SCM was
computed using $L = 12$ snapshots. The DL level was set to keep the
mean WNG for UC MVDR and DL MVDR beamformer equal. Over the observed
power output range, the UC MVDR beamformer has higher probability of
achieving lower output power compared to both SMI MVDR and DL MVDR
beamformers. The median output power of UC MVDR was more than 20 times
lower than SMI MVDR and about 10 times lower than DL MVDR
beamformer. Thus the UC MVDR suppresses the interferer better than
both SMI MVDR and DL MVDR beamformers. Moreover, the UC MVDR has
another advantage over the DL MVDR because it does not require an
\emph{a priori} choice of a tuning parameter like the DL factor.

\begin{figure}[!th]
  \centering
  \subfloat[]{\includegraphics[width=3in]{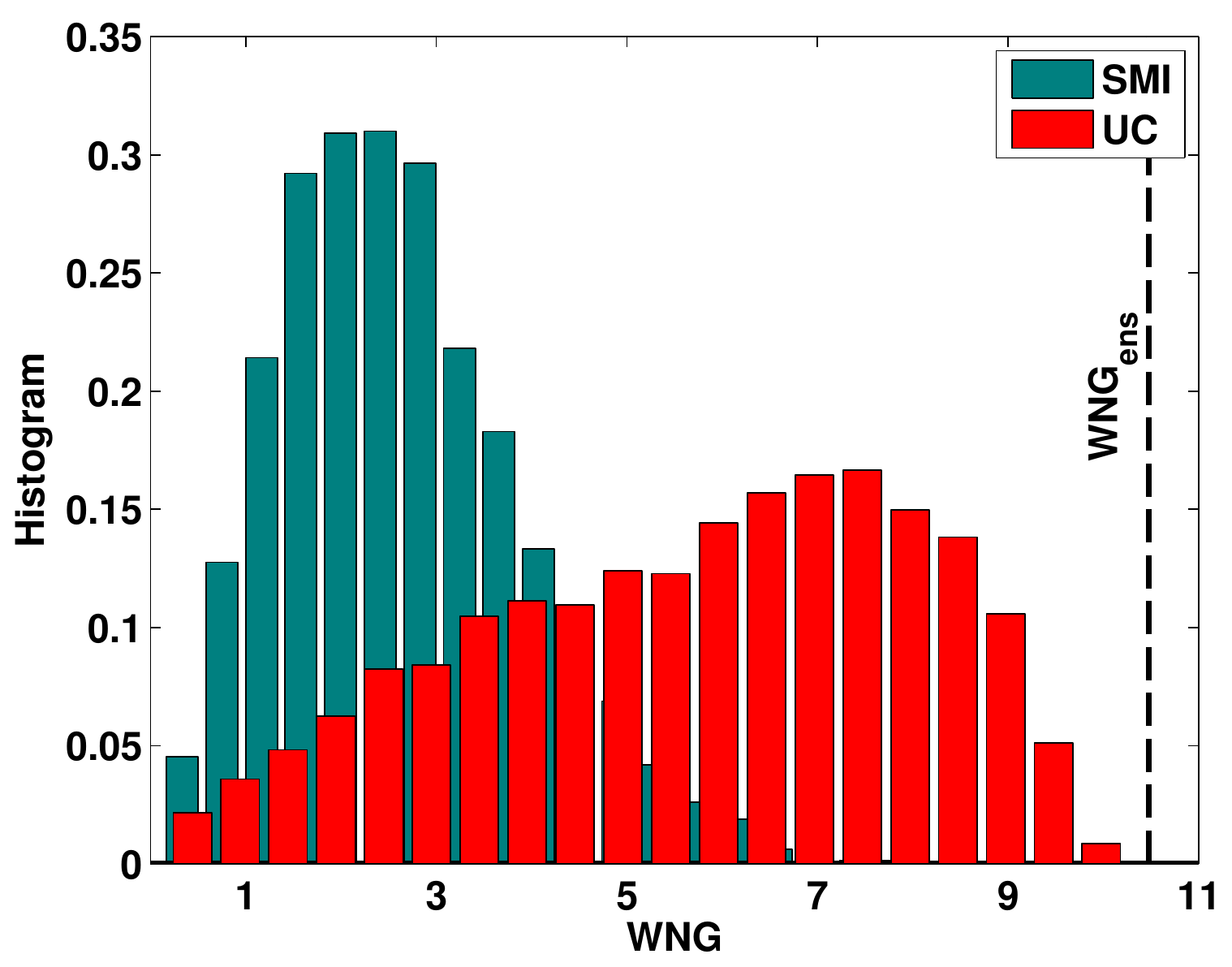}
  \label{fig:wng-hist-plot}}
  \hfill 
  \subfloat[]{\includegraphics[width=3in]{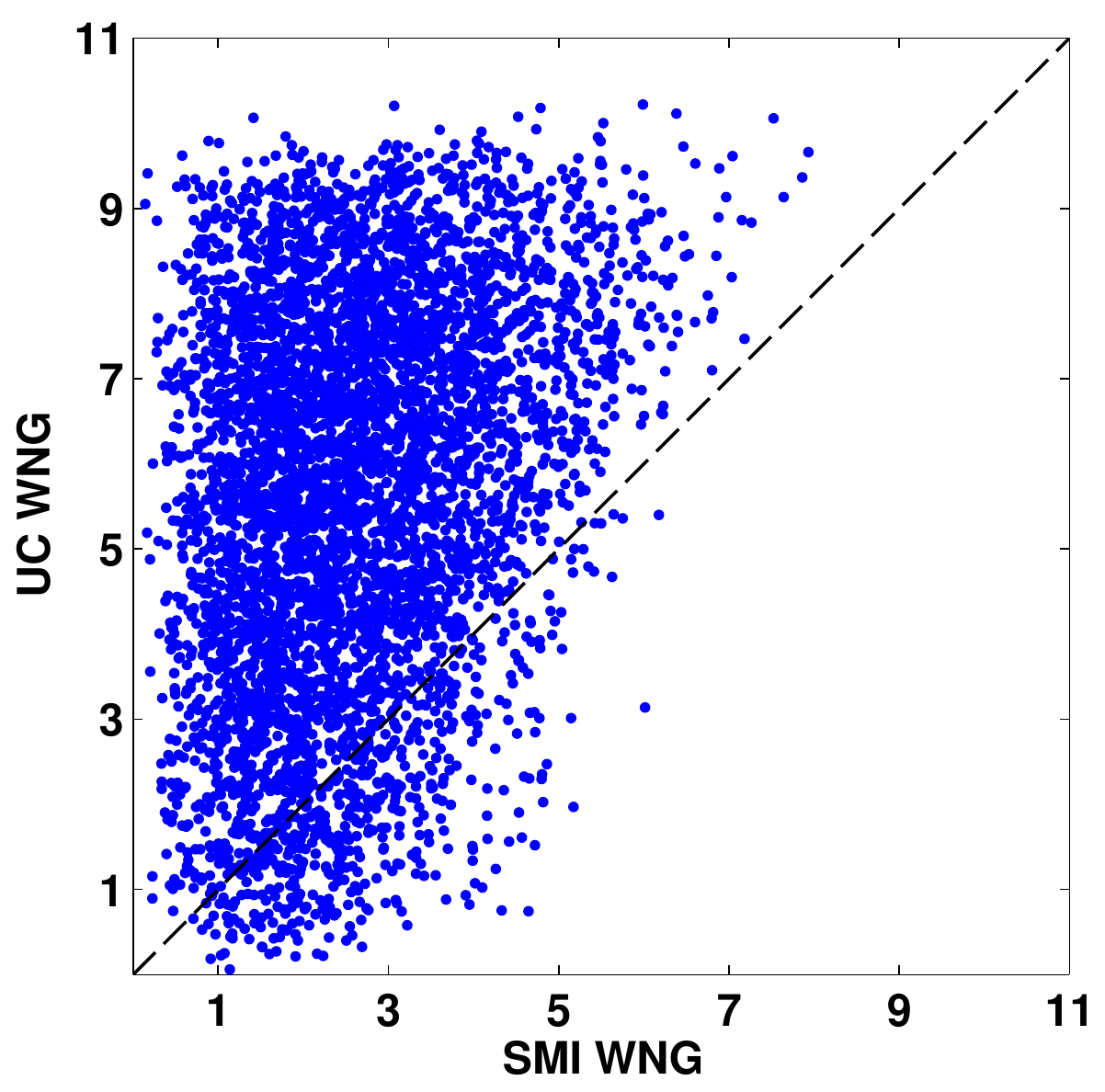}
    \label{fig:wng-scatter-plot}}
  \caption{Comparison between WNG for UC MVDR and SMI MVDR beamformer with
    $N = 11$ element ULA and $L = 12$ snapshots.}
  \label{fig:wng}
\end{figure}

\figurename{} \ref{fig:wng} compares the WNG for the UC MVDR against the SMI
MVDR beamformer for same Monte Carlo experiment used to generate
\figurename{} \ref{fig:ecdf-plots}. The optimal WNG for the experiment
is 11. The histograms in \figurename{} \ref{fig:wng-hist-plot} show
the improvement in WNG using the UC MVDR compared to SMI MVDR
beamformer for same set of data. The dashed vertical denotes the
ensemble WNG of 10.473. The UC MVDR beamformer has a higher
probability of achieving higher WNG with an average WNG of 5.672
compared to an average WNG of 2.629 using the SMI MVDR beamformer. The
scatter plot in \ref{fig:wng-scatter-plot} shows that the UC MVDR has
a higher WNG than SMI MVDR beamformer in each trial instance except
for small number of cases (bottom left corner in \figurename{}
\ref{fig:wng-scatter-plot}) where both beamformers have low WNG.

\section{Conclusion}
\label{sec:conclusion}
This paper presents the UC MVDR beamformer derived by moving the SMI
MVDR zeros to lie on the unit circle. By placing zeros on the unit
circle, the UC MVDR beampattern has perfect notches and lower
sidelobes when compared to the SMI MVDR beampattern. Numerical
simulations show that the UC MVDR beamformer suppresses interferers
better than the SMI MVDR and DL MVDR beamformers and has higher
average WNG than the SMI MVDR beamformer for the single interferer
case

%\vfill\pagebreak
%\section{REFERENCES}
%\label{sec:refs}
\bibliographystyle{IEEEtran}
\bibliography{IEEEabrv,myrefs}

\end{document}